\documentclass[twocolumn,prd,floatfix,preprintnumbers,a4paper,nofootinbib,superscriptaddress]{revtex4-1}

\usepackage{color}
\usepackage{calc}
\usepackage{amsmath,amssymb,graphicx}
\usepackage{amssymb,amsmath}
\usepackage{tensor}
\usepackage{bm}
\usepackage{microtype}
\usepackage{booktabs}
\usepackage{times}
\usepackage[varg]{txfonts}
\usepackage[colorlinks, pdfborder={0 0 0}]{hyperref}
\usepackage[caption=false]{subfig}
\definecolor{LinkColor}{rgb}{0.75, 0, 0}
\definecolor{CiteColor}{rgb}{0, 0.5, 0.5}
\definecolor{UrlColor}{rgb}{0, 0, 0.75}
\hypersetup{linkcolor=LinkColor}
\hypersetup{citecolor=CiteColor}
\hypersetup{urlcolor=UrlColor}
\allowdisplaybreaks
\usepackage[utf8]{inputenc}
\usepackage{ulem}
\newcommand{\MPl}{M_\text{Pl}}
\newcommand{\mll}{m_\textsc{ll}}
\newcommand{\lambdaul}{\Lambda_\textsc{ul}}

\newcommand{\<}{\begin{equation}}
\newcommand{\?}{\end{equation}}

\definecolor{mgreen}{rgb}{0.1,0.7,0.1}
\definecolor{lgreen}{rgb}{0.1,0.3,0.1}

\begin{document}

\title{Constraining black hole mimickers with gravitational wave observations}

\author{Nathan K. Johnson-McDaniel$^*$}
\affiliation{International Centre for Theoretical Sciences, Tata Institute of Fundamental Research, Bengaluru 560089, India}
\affiliation{Department of Applied Mathematics and Theoretical Physics, Centre for Mathematical Sciences, University of Cambridge, Cambridge, CB3 0WA, United Kingdom}
\altaffiliation{Current address: 149 Oak Avenue, Jefferson, Georgia 30549, USA}
\author{Arunava Mukherjee}
\affiliation{International Centre for Theoretical Sciences, Tata Institute of Fundamental Research, Bengaluru 560089, India}
\affiliation{Astroparticle Physics and Cosmology Division, Saha Institute of Nuclear Physics, 1/AF Bidhannagar, Kolkata-700064, India}
\affiliation{SUPA, School of Physics and Astronomy, University of Glasgow, Glasgow G12 8QQ, United Kingdom}
\affiliation{Albert-Einstein-Institut, Max-Planck-Institut f{\"u}r Gravitationsphysik, Callinstr.\ 38, D-30167 Hannover, Germany}
\affiliation{Leibniz Universit{\"a}t Hannover, D-30167 Hannover, Germany}
\author{Rahul Kashyap}
\affiliation{International Centre for Theoretical Sciences, Tata Institute of Fundamental Research, Bengaluru 560089, India}
\affiliation{Institute for Gravitation and the Cosmos, Center for Multimessenger Astrophysics, Department of Physics, The Pennsylvania State University, University Park, Pennsylvania 16802, USA}
\author{Parameswaran~Ajith}
\affiliation{International Centre for Theoretical Sciences, Tata Institute of Fundamental Research, Bengaluru 560089, India}
\affiliation{Canadian Institute for Advanced Research, CIFAR Azrieli Global Scholar, MaRS Centre, West Tower, 661 University Ave., Suite 505, Toronto, ON M5G 1M1, Canada}
\author{Walter~Del~Pozzo}
\affiliation{School of Physics and Astronomy, University of Birmingham, Edgbaston, Birmingham, B15 2TT, United Kingdom}
\affiliation{Universit{\'a} di Pisa, I-56127 Pisa, Italy}
\affiliation{INFN, Sezione di Pisa, I-56127 Pisa, Italy}
\author{Salvatore Vitale}
\affiliation{LIGO Laboratory and Kavli Institute for Astrophysics and Space Research, Massachusetts Institute of Technology, Cambridge, Massachusetts 02139, USA}


\begin{abstract}

LIGO and Virgo have recently observed a number of gravitational wave (GW) signals that are 
fully consistent with being emitted by binary black holes described by general relativity. 
However, there are theoretical proposals of exotic objects that can be massive and compact 
enough to be easily confused with black holes. Nevertheless, these objects differ from 
black holes in having nonzero tidal deformabilities, which can allow one to distinguish 
binaries containing such objects from binary black holes using GW observations. Using full 
Bayesian parameter estimation, we investigate the possibility of constraining the parameter 
space of such ``black hole mimickers'' with upcoming GW observations. Employing perfect 
fluid stars with a polytropic equation of state as a simple model that can encompass a 
variety of possible black hole mimickers, we show how the observed masses and tidal 
deformabilities of a binary constrain the equation of state. We also show how such 
constraints can be used to rule out binaries of some simple models of boson stars as 
possible sources of the simulated events we consider.

\end{abstract}
\maketitle 

\preprint{}
\date{\today}

\section{Introduction}

The Advanced LIGO~\cite{AdvLIGO} and Advanced Virgo~\cite{AdvVirgo} detectors have 
recently observed a number of gravitational wave (GW) signals from the coalescences of compact 
binaries~\cite{GW150914,GW151226,O1_BBH,GW170104,GW170814,GW170608,GW170817,GWTC-1,GW190425,GW190412,GW190814,GW190521_discovery}. The measured 
component masses and general characteristics of chirping signals in these events establish 
that the component objects in these binaries are extremely compact and massive, strongly 
suggesting that they are stellar-mass black holes~\cite{GW150914,GW151226,O1_BBH,GW170104,
GW170814,GW170608,GWTC-1,GW190412,GW190521_implications} except for two events that are likely to harbor neutron stars~\cite{GW170817,GW190425}, and one that could possibly
contain a neutron star~\cite{GW190814}. 
In addition, for the putative binary black hole events, the data are found to be fully 
consistent with binary black hole solutions in general relativity, as established by 
several consistency tests~\cite{GW150914_testingGR,GW151226,O1_BBH,GW170104,GW170814,GW170608,O2_TGR,GW190412,GW190814,GW190521_discovery,GW190521_implications}.

However, there are theoretical proposals of exotic alternatives to black holes, which can be 
massive  and compact enough so that GWs from binaries of such objects can be confused with 
those from binary black holes, e.g.,~\cite{Giudice:2016zpa, Mendes:2016vdr, 2016PhRvD..94h4031C, 
2017PhRvD..95h4014C, 2017NatAs...1..586C,Raposo:2018rjn}. One commonly considered alternative to black holes 
are boson stars~\cite{Liebling:2017fv}---gravitational equilibrium configurations of a 
massive, possibly self-interacting, scalar field; the axion is one possibility for such a 
field. Other alternatives include dark matter stars made of bosonic or fermionic particles 
(see, e.g.,~\cite{Giudice:2016zpa}). If the masses of these particles are sufficiently small 
or their self-interactions are sufficiently strong, they can form stars that are massive 
and compact enough to explain at least the general properties of the GW signals currently
 identified as coming from binary black hole coalescences. Other examples include 
gravastars---hypothetical objects with a de Sitter space interior surrounded by a shell of 
matter~\cite{Mazur:2004fk}.

Potential electromagnetic or astroparticle (e.g., neutrino) signatures of the coalescences 
of binaries containing such exotic objects are not well understood. Additionally, even in 
the case of scalar field stars (including boson stars), which is the most straightforward 
to model numerically, simulations of the coalescences of such binaries are still in the 
exploratory phase, e.g., \cite{2008PhRvD..77d4036P, 2016PhRvD..94h4031C,Bezares:2017mzk, 
Palenzuela:2017kcg, Helfer:2018vtq, Bezares:2018qwa} (see also~\cite{Sanchis-Gual:2018oui} 
for simulations of binaries of Proca stars and~\cite{CalderonBustillo:2020srq} for an analysis of
GW190521 with head-on Proca star binary simulations).
However, there are robust features that distinguish 
gravitational waveforms of binaries containing material objects from those of binary black 
holes. In particular, material objects will be deformed by the tidal field of their companion 
as well as by their own spin, and these deformations will affect the objects' gravitational 
fields, and thus the binary's gravitational waveform (from changes to its binding energy and 
radiative multipole moments, in the post-Newtonian picture). Post-Newtonian (PN) calculations 
of the effects of such changes to the multipole moments of the components of the binary are 
well-developed and can be used to model the inspiral waveforms of binaries of black hole 
mimickers, allowing one to distinguish binaries of such objects from binary black holes. 

The idea of constraining properties of specific black hole mimickers using constraints on 
the tidal deformabilities has been proposed in~\cite{2017PhRvD..95h4014C, Sennett:2017etc} 
(with initial work on constraining the tidal deformability of black holes in~\cite{Wade:2013hoa} 
and general theoretical proposals for such tests in~\cite{Porto:2016zng}). Additionally,
\cite{Krishnendu:2017shb,Krishnendu:2018nqa,Krishnendu:2019ebd,Krishnendu:2019tjp},
\cite{Maselli:2017cmm,Datta:2019epe}, and \cite{Cardoso:2019nis,Asali:2020wup} consider using
the spin-induced quadrupole moment, differences in tidal heating, and the resonant excitation 
of internal modes of the objects, respectively. All of the data analysis studies of using tidal deformabilities 
to constrain properties of black hole mimickers to date, including the recent one in~\cite{Pacilio:2020jza},
use the Fisher matrix approximation~\cite{Vallisneri:2007ev}, though there are recent full stochastic 
sampling studies of the constraints one can place using the spin-induced quadrupole 
moment~\cite{Krishnendu:2019tjp} and resonances~\cite{Asali:2020wup} as well as 
on constraints on tidal heating without considering black hole mimickers in 
particular~\cite{Datta:2020gem}. The Fisher matrix analyses are only a good 
approximation to the full analysis using stochastic sampling in the limit of very high signal-to-noise
ratio and when one can neglect the influence of non-Gaussian priors~\cite{Vallisneri:2007ev}. Thus,
one cannot make statistically robust statements using the Fisher matrix even for the loudest gravitational wave
events detected to date~\cite{GWTC-1}.

In this paper, we present a Bayesian method~\cite{Skilling2004a,Veitch:2009hd,LALInference} for distinguishing such exotic compact 
objects from binary black hole mergers based on the measurement of the tidal deformability. 
Alternatively, a null measurement will place an upper limit on the tidal deformability 
and hence will constrain the parameter space of various black-hole mimickers that can 
describe the signals being analyzed. This is the first full statistical analysis of this problem, 
without using the Fisher matrix approximation, and makes use of some of the best available 
gravitational waveforms. Additionally, it self-consistently takes into account the fact 
that black hole mimickers can merge at significantly lower frequencies than their black 
hole counterparts, both due to their larger radii and due to the tidal deformation that 
we aim to constrain.\footnote{Of course, one can confidently exclude binaries of black hole mimickers
with very large radii that merge well below the LIGO-Virgo band as sources of the observed signals.
However, the constraints we place here are on compact binaries that merge within the LIGO-Virgo band,
so placing constraints on these models using the observed frequencies (from, e.g., a spectrogram)
requires fairly refined, likely numerical, modeling of the binary of black hole mimickers, beyond the simple
computation of the contact frequency that we use. For instance, for an equal-mass binary of polytropic stars
modeling neutron stars, the peak frequency of the waveform is $\sim 2$ times that at contact---see Fig.~5
in~\cite{Bernuzzi:2012ci}.} We use perfect fluid stars with a polytropic equation of state as a 
simple model that can encompass a variety of black hole mimickers, though not all of them. In particular,
this does not include gravastars, with their negative tidal deformabilities.

We introduce the polytropic star model in Sec.~\ref{sec:poly_model} and the waveform model and contact frequency
calculation in Sec.~\ref{sec:waveform_model}. We analyze simulated observations in Sec.~\ref{sec:inj} and describe
the prospects for constraints on boson star models in Sec.~\ref{sec:boson_star_constraints}. Finally, we conclude in Sec.~\ref{sec:concl}.

\section{A simple model for non-spinning black hole mimickers}
\label{sec:poly_model}

For this initial study, we consider nonspinning perfect fluid stars described by a 
polytropic equation of state (EOS). The matter in the star is thus described by an energy 
density $\epsilon$ and pressure $p$, related by
\begin{equation}\label{eq:poly_EOS}
p = K \, (\epsilon - n \, p)^{1 + 1/n} = K \, \rho^{1 + 1/n},
\end{equation}
where $K > 0$ and $n > 0$ are the polytropic constant and index,
respectively, and $\rho = \epsilon - n \, p$ is the rest mass density. While this is a 
very simple model, polytropic EOSs nevertheless provide a good description of the structure 
of some potential black hole mimickers, as shown in~\cite{Giudice:2016zpa}. We also show 
that the tidal deformability-mass relationships for more compact boson stars 
(from~\cite{2017PhRvD..95h4014C,Cardoso_etal_data_URL}) are well approximated 
(at least in the quadrupolar and octupolar cases) by those of stars with polytropic EOSs. 
\begin{figure*}[tb]
	\includegraphics[width=\textwidth]{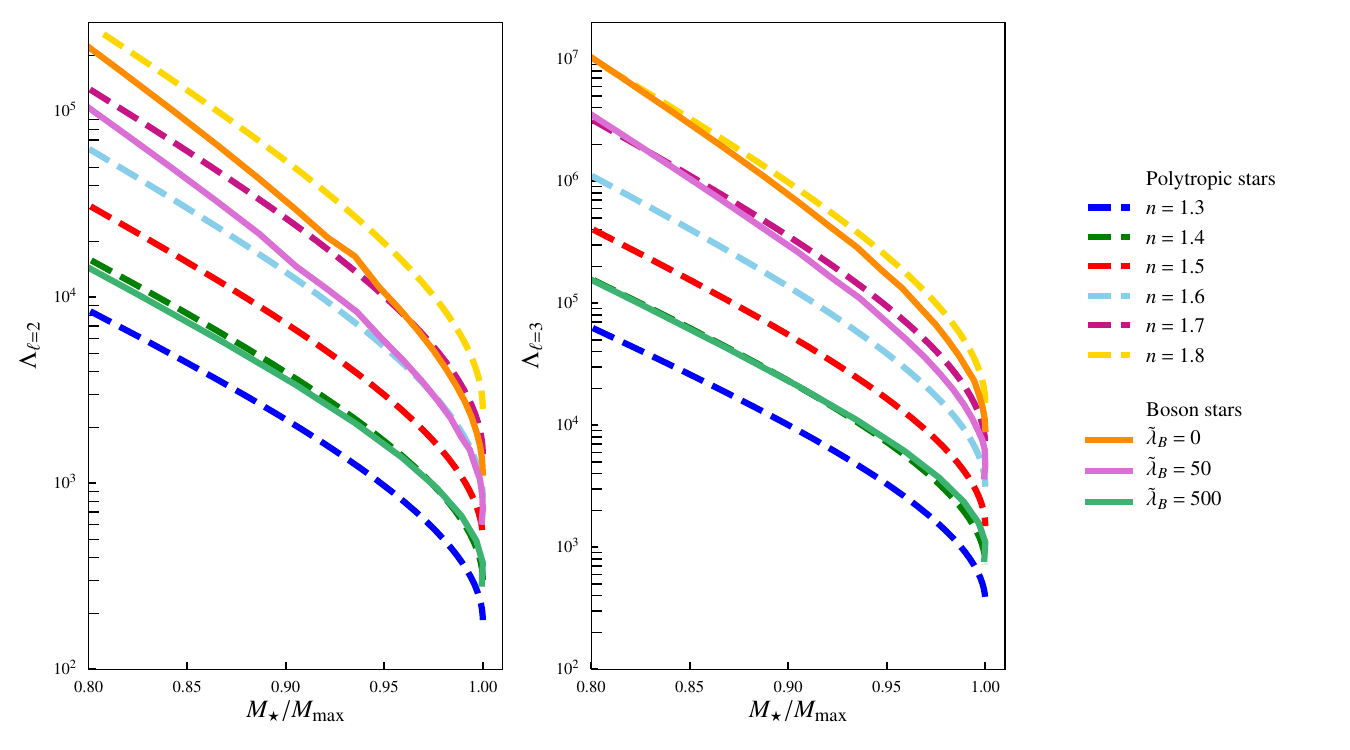}
	\caption{The quadrupolar and octupolar tidal deformabilities for polytropic stars as 
well as those for boson stars with a quartic self-interaction computed by~\cite{2017PhRvD..95h4014C}. 
Both of these are plotted versus the star's mass $M_\star$, scaled by the maximum mass of a 
stable star with a given EOS, $M_\text{max}$. Here $n$ is the polytropic index and 
$\tilde{\lambda}_B := \lambda_B(\MPl/m_B)^2$, where $m_B$ and $\lambda_B$ are the boson mass 
and coupling constant and $\MPl$ is the Planck mass.
	}
	\label{fig:polytropic_and_boson_star_Lambda_vs_M}
\end{figure*}

In Fig.~\ref{fig:polytropic_and_boson_star_Lambda_vs_M}, we compare the tidal deformabilities 
as a function of mass for boson stars with a quartic self-interaction from~\cite{2017PhRvD..95h4014C,Cardoso_etal_data_URL}
with those of polytropic stars of different polytropic index. In order to make these curves 
comparable, we scale the masses by the maximum mass. We find that the tidal deformabilities 
for larger values of $\tilde{\lambda}_B := \lambda_B(\MPl/m_B)^2$ are well approximated by 
those of polytropic stars. Here $m_B$ and $\lambda_B$ are the boson mass and coupling constant,
corresponding to a potential for the boson field $\phi$ of $V(|\phi|^2) = m_B^2 |\phi|^2 + \lambda_B |\phi|^{4}/2$,
and $\MPl$ is the Planck mass. We do not plot the results for the largest $\tilde{\lambda}_B$ or 
the solitonic boson star model from~\cite{2017PhRvD..95h4014C, Cardoso_etal_data_URL} as the 
data are not sufficiently finely sampled.

The properties of these stellar models that are relevant for this investigation are their 
radii and multipolar tidal deformabilities for a given mass. These quantities determine 
the contact frequency, as well as the effects of the objects' tidal interactions on the 
waveform. We only need the dominant quadrupolar tidal deformability to model the waveforms 
to the accuracy necessary for the present study, but require higher multipolar tidal 
deformabilities to accurately model the contact frequency. We compute the stellar structure 
and tidal deformabilities by solving the Oppenheimer--Volkov equations 
for the stellar structure and then using the expressions from~\cite{Landry:2014jka} to 
obtain the tidal deformabilities. The final outputs are the radii and tidal deformabilities 
as a function of mass up to the maximum mass allowed by the EOS.

Specifically, we consider the dimensionless quadrupolar tidal deformability 
$\Lambda := \lambda/M^5$, where $\lambda$ is the dimensionful quadrupolar tidal 
deformability and $M$ is the star's mass~\cite{Wade:2014vqa}. Here $\lambda$ is given 
by $Q_{ij} = -\lambda \, \mathcal{E}_{ij}$, where $Q_{ij}$ is the star's induced 
quadrupole moment and $\mathcal{E}_{ij}$ is the external tidal field (see, e.g., 
\cite{Flanagan:2007ix}), all measured in mass units. The minimum value of $\Lambda$ 
(for a stable star), $\Lambda_\text{min}$, is attained by the maximum mass ($M_\text{max}$) 
stable star, and since $\Lambda$ is dimensionless, $\Lambda_\text{min}$ only depends on 
$n$, not $K$. The $\Lambda_\text{min}$ values for the polytropic indices we consider range 
from $3.7$ for $n = 0.5$ to $8500$ for $n = 2$, and the associated minimum radii are 
$1.58$ to $6.75$ times the Schwarzschild radius 
associated with the star's mass.

\section{Inspiral waveforms for black hole mimickers}
\label{sec:waveform_model}

During the early stages of their inspiral, the compact objects can be approximated as 
point particles endowed with a tidally-induced quadrupole moment and their dynamics and 
gravitational waveforms can be computed in the PN approximation to general relativity, 
where one expands in terms of a small velocity parameter $v$ in units of 
$c$~\cite{Blanchet:2013haa}. The PN waveforms describing binaries of nonspinning point 
particles are currently known up to $\mathcal{O}(v^7)$ beyond the leading order in the phase.
Tidal effects first appear in the waveform as a high order $\mathcal{O}(v^{10})$ correction, 
but they can still have an appreciable effect on the waveforms depending on the value 
of the tidal deformability $\Lambda$.

The PN description of the point-particle portion of the waveform is not sufficiently 
accurate to describe the late inspiral where tidal effects are the largest (see, 
e.g.,~\cite{Buonanno:2009zt}). Thus, in the absence of accurate numerical simulations 
of binaries of black hole mimickers, we model the waveforms of these binaries of 
polytropic stars using as a base the frequency-domain binary black hole waveform 
model IMRPhenomD~\cite{Khan:2015jqa} (as implemented in LALSimulation~\cite{LALSuite}), 
to give a model for the point-particle part of the waveform that is more accurate in 
the late inspiral than pure PN results. In particular, IMRPhenomD contains effective 
$\mathcal{O}(v^9)$ though $\mathcal{O}(v^{11})$ terms in the inspiral phasing that are 
calibrated to numerical relativity simulations of binary black holes. To model the tidal 
effects, we add tidal corrections accurate to $\mathcal{O}(v^{12})$~\cite{Vines:2011ud} 
computed in the stationary phase approximation to these waveforms' phase, similar 
to~\cite{Barkett:2015wia}.\footnote{Since we add the tidal corrections to binary black 
hole waveforms, it is consistent for us to take the tidal Love number of a black hole 
to be zero, even with the ambiguity in the mapping from the standard calculations of 
Love numbers to the tidal terms in the waveform discussed by Gralla~\cite{Gralla:2017djj}.} 
The specific expression we use is Eq.~(A26) in~\cite{Wade:2014vqa}. We have checked that 
the dephasing due to the partially known higher order tidal terms and differences in 
tidal heating that we neglect here are small and/or do not improve 
agreement with numerical simulations of binary neutron stars. 

Specifically, we checked that all but one of various higher-order effects indeed lead to a negligible dephasing for the parameters we consider. In particular, we considered: i) the effects of the relative $1.5$PN [i.e., $\mathcal{O}(v^{13})$ relative to Newtonian] tidal tail term [Eqs.~(B8)--(B10) and~(B12) in~\cite{2012PhRvD..85l3007D}~]; ii) the leading $\mathcal{O}(v^{14})$ effects of the electric octupole tidal deformations [Eq.~(91) in~\cite{Yagi:2013sva}~]; iii) the leading $\mathcal{O}(v^{12})$ magnetic quadrupole tidal deformations [Eq.~(6) in the erratum to~\cite{Yagi:2013sva} using the quasi-universal relation between the electric and magnetic quadrupole tidal deformabilities for irrotational stars from Eq.~(31) and Table~IX of~\cite{Gagnon-Bischoff:2017tnz} to give an order-of-magnitude estimate];\footnote{The fit from \cite{Gagnon-Bischoff:2017tnz} gives the same order of magnitude and sign of the magnetic quadrupole tidal deformability found for boson stars in~\cite{2017PhRvD..95h4014C}. See Eqs.~(6) and~(20c) in~\cite{Landry:2018bil} for the phase in terms of the magnetic quadrupole deformation quantity used in~\cite{Gagnon-Bischoff:2017tnz}.} iv) the leading order $\mathcal{O}(v^{20})$ correction due to linearizing in the quadrupole tidal deformability [Eq.~(A5) in~\cite{Hinderer:2009ca}, with the corrections from the end of Appendix~B in~\cite{Yagi:2013sva}~]; and v) the leading $\mathcal{O}(v^{8})$ effects of horizon absorption in a binary of nonspinning black holes [calculated starting from Eq.~(11) in~\cite{Alvi:2001mx}~]. Horizon absorption effects will be present in the waveform model, due to its calibration to numerical relativity simulations of binary black holes. However, the tidal heating effects for binaries of black hole mimickers are expected to be much smaller than those for binaries of black holes, since the viscosities of black hole mimickers are expected to be much smaller than the effective viscosity of a black hole's horizon (see, e.g., the discussion for neutron stars in~\cite{Datta:2020gem}).

The dominant contribution is due to the $1.5$PN tail term and is less than $1.5$~rad for the cases we are considering (at the $90\%$ credible level of the posteriors that we estimate). We do not use the $1.5$PN phasing, since it does not agree as well with binary neutron star simulations as the $1$PN phasing (see Fig.~10 in~\cite{Dietrich:2018uni}), and the $2$PN phasing was not known completely when we started this project, though it has been obtained very recently~\cite{Henry:2020ski}. The dephasing from all the other contributions is at least an order of magnitude smaller, with the electric octupole tidal deformability giving the largest contribution, where the upper bound at the $90\%$ credible level is $< 0.11$~rad. For comparison, the largest $90\%$ bounds from the $0$PN and $1$PN tidal contributions are $6.1$ and $2.7$~rad, respectively.

Since our waveform models include the effects of tidal deformations using linear adiabatic 
tides, they are only accurate when the stars are sufficiently well-separated and not too 
tidally deformed. In particular, the models will definitely be inaccurate once the two 
stars have come into contact. We thus need to estimate the frequency at which a binary with 
given masses and tidal deformabilities will come into contact or have a tidal deformation 
greater than a given amount. Here we measure the tidal deformability using the star's 
fractional surface deformation $\Delta R/R$, computed as described below, and take the
maximum allowed value of $\Delta R/R$ to be $0.2$, at which point 
we assume that linear tidal deformations are no longer an accurate description of the 
system.\footnote{The value of $0.2$ comes from rough considerations of the size of the 
remainder from the approximation of linear tides, which should scale like $(\Delta R/R)^2$.
We want this to be smaller than unity, even if there is a reasonably large 
prefactor. However, as seen in Sec.~\ref{ssec:nonzero_lambda_inj}, this value will likely need
to be reduced to appropriately exclude the effects of waveform systematics.}

\subsection{Calculation of the contact frequency}
\label{sec:contact_freq_calc}
We estimate the contact frequency by extending the implicit expression for the contact 
separation given for binary neutron stars in Eq.~(77) of~\cite{Damour:2009wj}. This expression 
includes the leading effects of tidal deformability on the contact separation using 
the stars' quadrupolar shape tidal deformabilities~\cite{Damour:2009vw,Landry:2014jka}. 
We extend this expression to include higher multipolar contributions to the tidal deformabilitie and higher PN corrections to the 
tidal fields, from~\cite{JohnsonMcDaniel:2009dq}. We then compute the frequency from the 
contact separation using a PN expression. When we apply this expression to binary neutron 
stars, we find good agreement with numerical relativity results for polytropic stars from 
the literature~\cite{Bernuzzi:2012ci, Radice:2013hxh,Dietrich_PC}. 
In this section, we use geometrized units with $G = c = 1$.

We compute the separation between the stars when their distorted surfaces come into contact, including the multipolar shape deformations through $\ell = 5$ and the $1$PN corrections to the quadrupolar and octupolar tidal fields, by solving
\begin{widetext}
\begin{equation}\label{eq:contact_sep}
	R_c = \left[1 + h_2^{(1)}\frac{m_2}{m_1}\left(1 - \frac{m_2}{2R_c} - \frac{M}{R_c}\right)\left(\frac{R_1}{R_c}\right)^3 + h_3^{(1)}\frac{m_2}{m_1}\left(1 - \frac{3m_2}{R_c} - \frac{M}{2R_c}\right)\left (\frac{R_1}{R_c}\right)^4 + h_4^{(1)}\frac{m_2}{m_1}\left(\frac{R_1}{R_c}\right)^5  + h_5^{(1)}\frac{m_2}{m_1}\left(\frac{R_1}{R_c}\right)^6 \right]R_1 + (1 \leftrightarrow 2)\\
\end{equation}
\end{widetext}
for the contact separation $R_c$, where $m_A$ and $R_A$ are the stars' masses and unperturbed radii, respectively, $M := m_1 + m_2$ is the binary's total mass, and $h_k^{(A)}$ is the $k$th shape (or surficial) Love number for star $A$~\cite{Damour:2009vw,Landry:2014jka}. The notation ``$+ (1 \leftrightarrow 2)$'' implies that the preceding expression is added to itself with all the $1$ and $2$ star labels swapped (so this operation does not act on the ``2'' subscript in $h_2^{(A)}$). We compute each star's shape Love numbers and radius using its mass and the specific polytropic EOS we are considering. The expression for all $\ell$ to Newtonian order is $R_c = [1 + (m_2/m_1)\sum_{\ell = 2}^\infty h_\ell^{(1)}(R_1/R_c)^{\ell+1}]R_1 + (1 \leftrightarrow 2)$, using the Newtonian potential of the star's companion at a distance of $R_c$ to compute the electric tidal fields in Eq.~(2.2) of~\cite{Landry:2014jka} and taking the unit vector $\Omega_k$ in that paper's Eq. (2.6) to point towards the star's companion. The $1$PN corrections to the quadrupole and octupole tidal fields come from Eqs.~(B1a) and~(B1c) in~\cite{JohnsonMcDaniel:2009dq} evaluated at $t = 0$, noting that here we take $\Omega_k = -\hat{x}_k$ for the vector that points from star $1$ to star $2$ at $t = 0$. The expression in square brackets in Eq.~\eqref{eq:contact_sep} is the fractional surface deformation of star~1.

We then solve this equation numerically for $R_c$, noting that there is a unique positive root for all contact radii for which we can possibly trust the post-Newtonian expressions used to obtain it. This follows because the shape Love numbers are all positive, so the right-hand side of the equation will be a decreasing function of $R_c$ (for positive $R_c$) provided that the two terms from the $1$PN corrections to the quadrupole and octupole tidal fields are positive. Since the left-hand side is an increasing function of $R_c$ (for positive $R_c$), there can thus only be one root. For the $1$PN corrections to the quadrupole and octupole tidal fields to be positive, it is sufficient to assume that $R_c > 3.5M$, and one would not want to consider much smaller separations using the post-Newtonian approximation. 

We convert $R_c$ to $f_c$, the binary's (dominant quadrupole mode) gravitational wave frequency at contact, using the $3$PN point-particle relation, augmented with the tidal corrections. Specifically, we take the $3$PN point particle relation between $M/R_c$ and $x_c$ [e.g., Eq.~(231) in~\cite{Blanchet:2013haa}~] and expanding its reciprocal to $3$PN. Here we take the gauge constant $r_0' = M$ and use the Newtonian relation between $x_c$ and $M/R_c$ inside the logarithm. We use the expanded version of the reciprocal since this removes the $\eta$-independent terms at $2$ and $3$PN in the series. We then add on the quadrupole tidal pieces through $1$PN from Eq.~(2.9) in~\cite{Vines:2011ud}, and the higher-order tidal pieces through $\ell = 5$ to Newtonian order from Eq.~(A6) in~\cite{Yagi:2013sva}. All of these contributions are linearized in the Love number. This gives
\begin{widetext}
\begin{align}
\frac{R_c}{M} & = \frac{1}{x_c} - 1 + \frac{\eta}{3} + \left(\frac{19}{4} + \frac{\eta}{9}\right)\eta x_c + \Biggl(-\frac{24257}{2520} + \frac{41}{192}\pi^2 - \frac{37}{12}\eta + \frac{2}{81}\eta^3 + \frac{22}{3}\log x_c\Biggr) \eta x_c^2 \\ \nonumber 
& + \frac{2}{3}\left(\frac{m_2}{m_1}\Biggl\{\Biggl[3 - \left(3 - 13\frac{m_2}{M} + \frac{m_2^2}{2M^2}\right)x_c\Biggr]\tilde{\lambda}^{(1)}_2 + 4\tilde{\lambda}_3^{(1)}x_c^2 + 5\tilde{\lambda}_4^{(1)}x_c^4 + 6\tilde{\lambda}_5^{(1)}x_c^6\Biggr\} + (1\leftrightarrow 2)\right)x_c^4,
\end{align}
\end{widetext}
where $x_c = (\pi M f_c)^{2/3}$ is the standard PN parameter (evaluated at the contact frequency), $\eta := m_1m_2/M^2$ is the binary's symmetric mass ratio and $\tilde{\lambda}^{(A)}_\ell = k^{(A)}_\ell(R_A/M)^{2\ell + 1}$ denotes a quantity related to the $\ell$th electric tidal Love number of the $A$th star, scaled by $M^{2\ell+1}$ [cf.\ Eq.~(9) in~\cite{Yagi:2013sva}, but note that Yagi's $\bar{\lambda}_\ell$ is scaled using the individual star's mass, while we scale using the total mass of the binary, and do not include the usual numerical factors present in the definition of the Love number]. We solve this equation numerically. [The leading-order tidal contribution inside the curly brackets is $\sum_{\ell = 2}^\infty (\ell + 1)\tilde{\lambda}_\ell^{(1)}x_c^{2\ell-4}$ to all orders.]

We have not attempted to account for the difference in the Schwarzschild coordinate system used for the perturbed compact objects and the post-Newtonian harmonic coordinate system used for the binary. It is possible to estimate the effects of this coordinate transformation in the post-Newtonian approximation, using the binary black hole asymptotic matching results from~\cite{JohnsonMcDaniel:2009dq} (though bearing in mind that the asymptotic matching calculation will be pushed to---and quite possibly beyond---the limits of its validity at the point of contact). Here, the leading-order effect is given by $r_\text{PN} = r_\text{Schw} - M_\star$, where $r_\text{PN, Schw}$ denotes the radial coordinate in one of the two coordinate systems and $M_\star$ is the  mass of the compact object the Schwarzschild coordinates are describing. (One would apply this transformation to the perturbed radii of the stars when computing the contact separation, which is given in PN harmonic coordinates.) This transformation thus increases the inferred contact frequency.

However, as in the initial calculation by~\cite{Damour:2009wj}, we do not use this transformation here: We found that the expression without the transformation reproduces the numerical relativity (NR) estimates of the contact frequency quite well, at least in the equal-mass $n = 1$ polytropic cases for which they are available. Specifically, \cite{Bernuzzi:2012ci} quotes a dimensionless (dominant quadrupole mode) contact frequency of $Mf_c \simeq 0.078$ for a binary of equal-mass $n=1$ polytropic stars with compactnesses of $0.16$, while \cite{Radice:2013hxh} reports a dimensionless contact frequency of $M f_c \simeq 0.11$ for a binary of such stars with compactness $0.18$. (Here $M$ denotes the sum of the gravitational masses of the stars in isolation.) Our expression gives dimensionless contact frequencies $M f_c$ of $0.0778$ and $0.111$ for those two systems.\footnote{These values have estimated fractional errors of $\lesssim 2\%$ each from the truncation of the PN expansion in the tidal fields or the radius-to-frequency conversion, and $\lesssim 1\%$ from the truncation in the multipole expansion of the tidal deformation. These errors are estimated using the ratio of the next-to-highest and highest order results.} While it is possible that the true values of these contact frequencies are higher, as this quantity is quite sensitive to the specific isodensity contour used to compute it, and generally increases with resolution~\cite{Dietrich_PC}, using the lower frequencies given by the expression without including the coordinate transformation is more conservative. In the future, we will compare our waveform model with numerical relativity waveforms (starting with those for binary neutron stars) to ascertain if our estimate of the contact frequency is an accurate guide for when the waveform model becomes inaccurate---we have performed some of the first such tests in Sec.~\ref{ssec:nonzero_lambda_inj}. Note that in the following, for brevity, we will use the term ``contact frequency'' to refer to the minimum of this contact frequency and the frequency at which one of the objects first has a fractional surface deformation of $0.2$.

\section{Constraining the properties of polytropic stars with GW observations}
\label{sec:inj}

\begin{table}
\centering
\begin{tabular}{c@{\quad} c@{\quad}c@{\quad}c@{\quad}r}
\toprule
$m_1^z, m_2^z $ &  $d_L$  & SNR with the \\
$[M_\odot]$  & [Mpc] &  ``late-high'' PSD \\
\midrule
39.5, 31.7  & 397.7 & 43.5  \\
36.8, 22.9  & 592.4 & 24.7  \\
27.6, 15.6  & 620.6 & 18.2  \\
15.5, 8.2   & 292.5 & 24.4  \\
\bottomrule
\end{tabular}
\caption{Detector frame (i.e., redshifted) masses and luminosity distance $d_L$ of simulated 
binary black-hole events, replicating the first four GW signals observed by the LIGO 
detectors~\cite{GW150914, GW151226, O1_BBH, GW170104, GWTC-1}. We give the SNRs of the simulated signals 
in Advanced LIGO's ``late-high'' PSD~\cite{Aasi:2013wya} (comparable to the 
sensitivity in the third observing run O3).}
\label{tab:inj_events}
\end{table}

\begin{figure*}[tb]
	\includegraphics[width=0.6\textwidth]{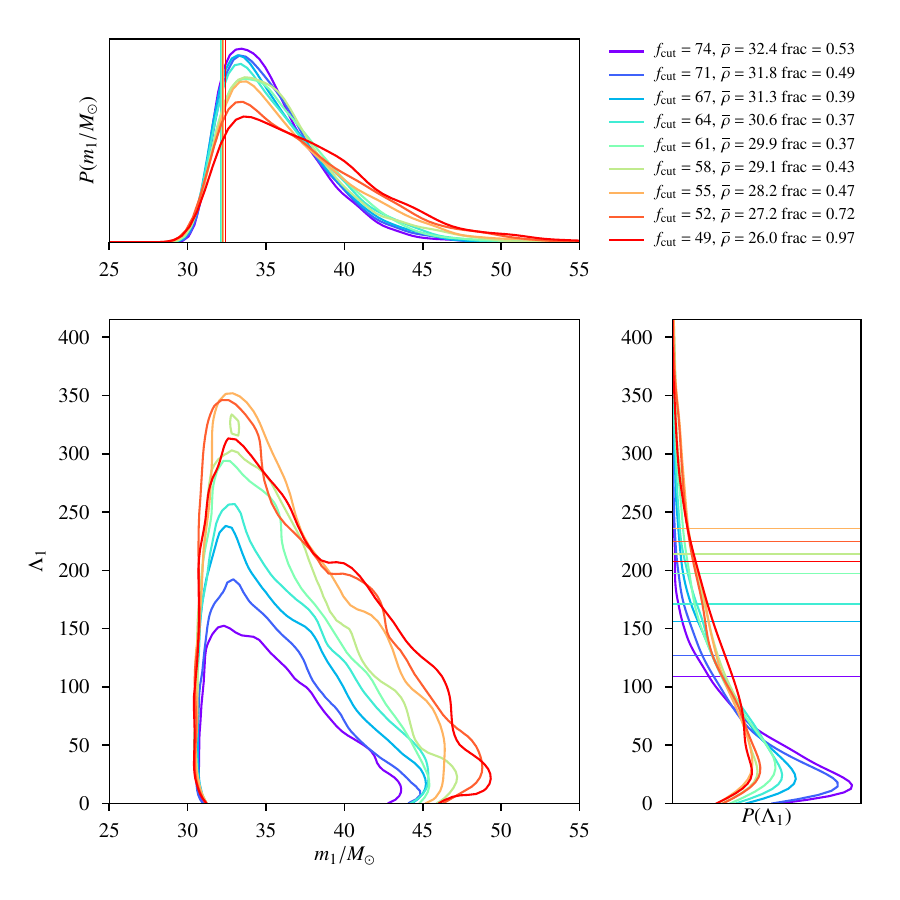}
	\caption{Posterior distributions of the (source frame) mass $m_1$ and tidal deformability $\Lambda_1$ 
of the more massive object (marginalized over all other parameters), obtained from the simulated 
observation with detector frame masses $m_1^z = 39.5 M_\odot$ and $m_2^z = 31.7 M_\odot$. The middle panel shows the 
90\% credible regions in the marginalized posterior distribution $P (m_1, \Lambda_1)$ while 
top/side panels show the marginalized one-dimensional posteriors $P(m_1)$ and $P(\Lambda_1)$. 
The distributions are computed from the posterior samples using Gaussian kernel density 
estimates. The legend shows the cutoff frequencies employed in the calculation of these 
posteriors, the mean SNR of the posterior samples, and the fraction of posterior samples with 
contact frequency (computed using $n = 0.9$) larger than the cutoff frequency employed. The 
vertical lines on the top panel show the 90\% credible lower bounds on $m_1$ while the 
horizontal lines on the side panel show the 90\% credible upper bounds on $\Lambda_1$.  
	}
	\label{fig:posterior_plots}
\end{figure*}

\begin{figure*}[tb]
\includegraphics[width=\textwidth]{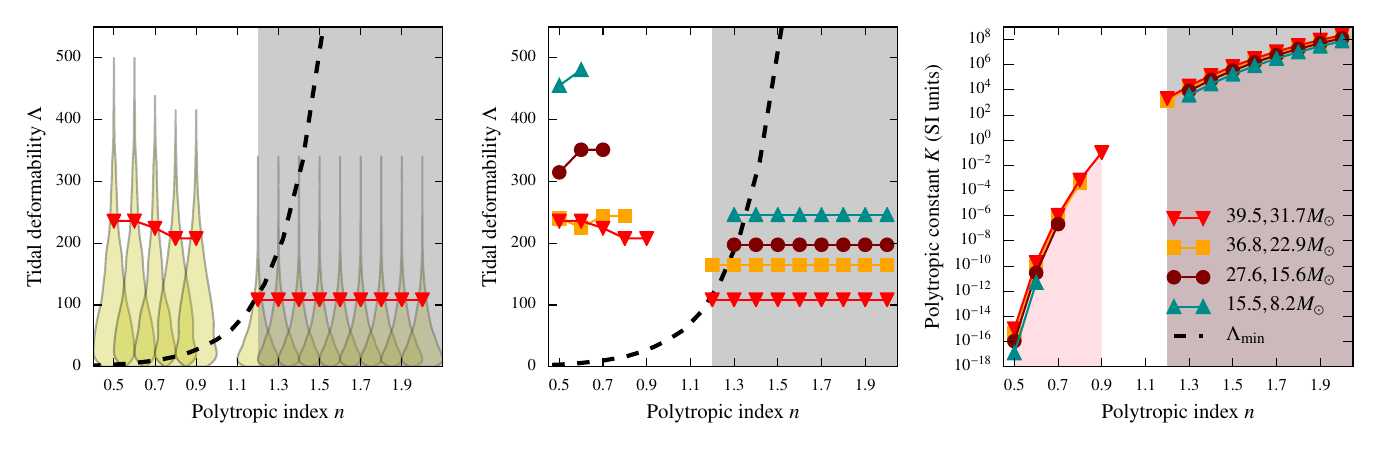}
\caption{\emph{Left panel:} The dashed line shows the minimum  value $\Lambda_\mathrm{min}$ 
of the dimensionless tidal deformability parameter allowed by the polytropic EOS [Eq.~\eqref{eq:poly_EOS}],
plotted against the polytropic index $n$. The violin plots (yellow) show the marginalized posterior
distribution of $\Lambda_1$ for each value of $n$ (thus the individual violin plots' extent 
in the horizontal direction indicates probability, on an arbitrary scale, not values of $n$), 
while the red triangles show the 90\% credible upper bounds $\lambdaul$. These are both 
computed from a simulated event with (detector frame) masses of $39.5$ and $31.7 M_\odot$. 
The shaded region shows the values of $n$ for which the observed upper bounds $\lambdaul$ 
are lower than the theoretical minimum  $\Lambda_\mathrm{min}$ for the EOS, and hence can 
be ruled out. In all panels, the values of $n$ for which there are no points for a given 
event are those values for which there are not enough posterior samples 
satisfying the selection criteria discussed in the text (i.e., contact frequency $>$ 
cutoff frequency used in the parameter estimation or $\Lambda_1 < \Lambda_\text{min}$). 
\emph{Middle panel:} Same as the left panel, except that we show the upper bounds obtained 
from all four simulated events (masses shown in legend). We show the largest exclusion 
region, corresponding to the most massive system considered (the same one as in the left 
panel). \emph{Right panel:} The maximum value of the polytropic constant $K$ (for a given 
polytropic index $n$) that is unconstrained by the observed lower limit $\mll$ on the mass 
of the compact object for all four simulated events. The pink shaded region is the largest 
region excluded by such an observation, and corresponds to the most massive system 
considered (the same one as in the left panel). The gray shaded region corresponds to 
the $\Lambda_\mathrm{min}$ constraint from the same system.}
\label{fig:polytrope_constraints}
\end{figure*}

GW signals from the quasi-circular inspiral of nonspinning compact binaries are characterized 
by the set of intrinsic parameters $\{m_1, m_2, \Lambda_1, \Lambda_2\}$ (as well as 
higher-order tidal deformability and tidal heating coefficients that are negligible at the 
accuracies to which we are working) and several extrinsic parameters describing the binary's location 
and orientation and the arrival time of the signals at the detector. We simulate observations of 
nonspinning binary black holes with parameters similar to LIGO's first four putative binary 
black hole events, using the median values of the masses in the detector frame (i.e., 
including cosmological redshift) $m_1^z, m_2^z$~\cite{GW150914, GW151226, O1_BBH, GW170104, GWTC-1}.\footnote{Note 
that all the LIGO-Virgo events that we simulate, except the lowest-mass one, are consistent 
with binaries of \emph{nonspinning} objects at the $90\%$ credible level~\cite{GW150914,
GW151226,O1_BBH,GW170104,GWTC-1}.} We use these events as proxies for future detections, but this
selection does not model a sample of detections expected in future observing runs.

The sky location and orientation of the binaries were chosen in such a way that the 
signal-to-noise ratio (SNR) of the signals is consistent with the expected distributions in 
the upcoming observations (see Table~\ref{tab:inj_events}). We then estimate the posterior 
distributions of the extrinsic and intrinsic parameters describing the signals using the 
\textsc{LALInferenceNest} code~\cite{Veitch:2009hd}, which provides an implementation of 
the Nested Sampling algorithm~\cite{Skilling2004a} in the \textsc{LALInference} 
software~\cite{LALInference}. We use flat priors on $m_1^z, m_2^z, \Lambda_1, \Lambda_2$; for 
derived quantities we use the priors induced from these flat priors. We have checked that 
the data are sufficiently informative that the posteriors do not resemble the priors. We infer the source
frame masses (used, e.g., to compare with the maximum mass allowed by the equation of state) using the redshift obtained from the inferred luminosity distance using a standard
$\Lambda$CDM cosmology with the cosmological parameters given by the TT+lowP+lensing+ext values
from~\cite{Ade:2015xua}. The parameter estimation is performed assuming a two-detector Advanced LIGO network with 
sensitivity anticipated in the upcoming observational run [modeled by the ``late-high'' 
power spectral density (PSD)~\cite{Aasi:2013wya} with a low-frequency cutoff of $15$~Hz]. 
We choose to present results using the sensitivity similar to that in the third observing run (O3) since we find when analyzing simulated observations with
the observed SNRs of the events we use as models here (about half the SNRs with the O3-like sensitivity given in Table~\ref{tab:inj_events}) that we
are not able to derive interesting constraints using the self-consistent analysis that we describe below. We do not include any detector noise
in these simulations, in effect averaging over noise realizations~\cite{Nissanke:2009kt}, so
the only contribution from the detectors' noise is in the likelihood. Future work will consider the effects of realistic detector noise
as well as uncertainty in detector calibration~\cite{Sun:2020wke}.

Note that, depending on the EOS, the compact objects can come in contact at different GW 
frequencies, causing the inspiral to end. While we could truncate the waveforms (or equivalently, the likelihood integral) at the contact frequency for each set of parameters, this would bias the results, by favoring parameters that have a higher-frequency cutoff, and thus more signal included in the likelihood integral (see, e.g.,~\cite{Mandel:2014tca} for further discussion of the problems with using a parameter-dependent cutoff). Thus, for each simulated observation, we compute 
the posterior distributions by truncating the likelihood integral at different cutoff 
frequencies, and compute the 90\% credible upper limits on $\Lambda_1, \Lambda_2$ and lower 
limits on $m_1, m_2$ from the marginalized one-dimensional posteriors. When constraining 
the EOS, for each value of the polytropic index $n$, we choose the upper/lower limits 
derived using an appropriate choice of the cutoff frequency (see below). 

We compute the contact frequency assuming that both stars have the same polytropic index $n$, 
but possibly different polytropic constants $K$.\footnote{We allow the stars to have different polytropic constants 
since we do not expect the structure of black hole mimickers to be described exactly by a 
polytropic equation of state. We could also allow the stars to have different polytropic 
indices, but for simplicity have not allowed that extra freedom in this initial study.}
We necessarily restrict this computation to the posterior samples that have both $\Lambda$s above the minimum allowed 
by the EOS---the samples that do not satisfy this condition are included, as we cannot calculate the contact frequency in this
case. (We comment later on the effects of this choice.) We then choose the largest cutoff frequency 
such that 90\% of the posterior samples correspond to a contact frequency greater than the cutoff frequency employed in 
the posterior computation. If 
the best 90\% credible upper bound $\Lambda_\textsc{ul}$ on the tidal deformability is 
smaller than the $\Lambda_\mathrm{min}$ corresponding to this EOS, the data exclude the 
possibility of both stars being described by this EOS at high confidence. Moreover, for a 
given polytropic index $n$, the minimum $M_\mathrm{max}$ sets the minimum allowed value of 
the polytropic constant $K$. Therefore, the best 90\% credible lower bound $m_\textsc{ll}$ 
on the component masses puts a constraint on $K$ for the values of $n$ that are not ruled 
out by the tidal deformability constraint.

Figure~\ref{fig:posterior_plots} shows the posterior distributions of the mass $m_1$ and tidal deformability $\Lambda_1$ of the more massive object (marginalized over all other parameters), obtained from the simulated observation with $m_1^z = 39.5 M_\odot$ and $m_2^z = 31.7 M_\odot$. We show the cutoff frequencies employed in the calculation of these posteriors, assuming a polytropic index of $n = 0.9$, in the legend. The legend also shows the mean SNR $\bar{\rho}$ of the posterior samples---larger cutoff frequencies produce larger SNRs and hence narrower posteriors. The fraction of posterior samples with contact frequency (with $n = 0.9$) larger than the cutoff frequency employed is also shown in the legend. Here the largest cutoff frequency such that this fraction is larger than 0.9 is $f_\mathrm{cut} = 49$ Hz, from which we read off the 90\% credible upper limit on $\Lambda_1$ and the 90\% credible lower limit on $m_1$.\footnote{We include the samples with contact frequency less than the cutoff frequency when computing this upper limit. However, we have checked that if one excludes those samples, the largest difference in the upper limit on $\Lambda_1$ is only $8\%$ (for the lowest-mass case and $n = 0.5$) and in that case the version with all the samples gives a larger upper bound, and is thus more conservative. For the cases where the version with all the samples gives a smaller upper bound, the largest difference is only $1\%$. The difference in the 90\% credible lower bound on $m_1$ is at most $0.3\%$}

Figure~\ref{fig:polytrope_constraints} (left and middle panels) shows the estimated 
$\Lambda_\textsc{ul}$ for various polytropic indices. The shaded region corresponds to 
observed $\Lambda_\textsc{ul} < \Lambda_\mathrm{min}$ (theoretical), and is therefore 
ruled out by such a GW event. In the right panel, in addition to the excluded region 
in $n$ from the tidal deformation, we also show the shaded region in $n - K$ space that 
is ruled out by having the observed $m_\textsc{ll} > M_\mathrm{max}$.

We find that for the values of $n$ for which we have sufficient numbers of samples with 
$\Lambda$s above the minimum allowed by the EOS (i.e., just the ones to the left of the 
gap in the middle panel of Fig.~\ref{fig:polytrope_constraints}, with $n$ at most $0.9$), 
we constrain the radii of the stars (at the $90\%$ credible level) to be at most $2.6$--$3.5$ 
times the Schwarzschild radius associated with their mass (increasing with increasing $n$); 
these range from $1.4$ to $2.0$ times the minimum radius allowed by the EOS. The masses of 
the stars are constrained to be at least $60\%$ to $90\%$ of the maximum mass, with this fraction 
increasing with increasing $n$.

The upper cutoff frequencies used in our analysis range up to $1.2$ times the dominant 
gravitational wave frequency corresponding to the innermost stable circular orbit of a 
Schwarzschild black hole with the same mass as the binary's total detector-frame mass, 
$f_\text{ISCO} = [\pi 6^{3/2}(m_1^z + m_2^z)]^{-1}$ (in $G = c = 1$ units), in steps of 
$0.05f_\text{ISCO}$. Here $m_{1,2}^z$ are the detector-frame individual masses of the binary. The lower
cutoff frequency is always $15$~Hz. The results shown come from upper 
cutoff frequencies of $0.8$, $0.85$, or $0.9$ times $f_\text{ISCO}$ for the cases to the 
left of the gap in Fig.~\ref{fig:polytrope_constraints}, and $1.2f_\text{ISCO}$ for the 
ones to the right of the gap. This is why the constraints to the right of the gap have no $n$
dependence. The frequency of $1.2f_\text{ISCO}$ is still well below the 
binary black hole merger frequency (taken to be the frequency of the peak of the dominant 
mode of the waveform and computed using the fit from~\cite{Bohe:2016gbl}) for all samples 
we consider. These upper cutoff frequencies correspond to orbital separations of 
$\sim 200$--$700$~km for the systems considered.

We have checked that if instead of keeping all samples where at least one star has $\Lambda$ below
the minimum allowed by the EOS, we instead compute the contact frequency by setting $\Lambda = 0$ and
using a compactness of $0.5$ (as for a Schwarzschild black hole) for such stars, then we obtain a smaller
fraction of samples that have a contact frequency greater than the cutoff frequency in some cases. Here we
take the minimum of the contact frequency and the binary black hole peak frequency using the fit from~\cite{Bohe:2016gbl},
However, the binary black hole peak frequency is above the cutoff frequency in the cases we consider (including the
simulated observations with nonzero tidal deformabilities considered below), so this use of the peak frequency fit does not end up affecting any results.

For the results in Fig.~\ref{fig:polytrope_constraints}, using the Schwarzschild values when $\Lambda$ is below the minimum allowed by the EOS does not affect the fractions
of samples for small polytropic indices, below the gap in the plot. However, for the larger polytropic
indices, above the gap in the plot, this procedure significantly reduces the largest fraction to be smaller
than $90\%$ for the first two to three cases above the gap. For instance, in the GW150914-like case and
$n = 1.2$, it reduces the fraction at the highest cutoff frequency from $98\%$ to $67\%$, so the second
smallest cutoff frequency has the largest fraction, $77\%$. For $n = 1.3$ with this computation, the
smallest cutoff frequency gives the largest fraction, $95\%$. Nevertheless, for the larger polytropic indices ($n \geq 1.4$
in the GW150914-like case), the largest cutoff frequency has more than $90\%$ of samples with contact frequency greater than the cutoff frequency.
Thus, the results for those indices in Fig.~\ref{fig:polytrope_constraints} would be unchanged with this alternate computation
of the contact frequency.

\subsection{Simulated observations with nonzero tidal deformabilities}
\label{ssec:nonzero_lambda_inj}

We have also investigated how well this method performs when analyzing signals with nonzero tidal
deformability. Specifically, we have considered simulated observations of compact binary inspirals modeled by the IMRPhenomD
+  PN tides waveform model (the same model we use for parameter estimation). We also employ simulated observations of numerical
relativity waveforms of binary neutron stars (BNSs) scaled in frequency to give binary black hole-like masses. All of these simulated observations
use GW150914-like mass and distance parameters. For the IMRPhenomD + PN tides simulated observation we use the same detector frame masses of $39.5$ and $31.7M_\odot$
and distance of $397.7$~Mpc as for the binary black hole simulated observation, as well as $\Lambda_1 = \Lambda_2 = 200$ (choosing the same
values for both stars for simplicity, rather than because this is expected for any particular black hole mimicker model in such an unequal-mass case).
The BNS waveforms we consider are hybrids of the TEOBResumS effective-one-body waveform model~\cite{Nagar:2018zoe} and
numerical relativity simulations from the CoRe database~\cite{Dietrich:2018phi} with the 2B and ALF2 equations of state, the latter with
two different masses and thus different tidal deformabilities. These hybrids were used in~\cite{Chen:2020fzm}, which gives further details
about the simulations and construction of the hybrids. These are all equal-mass systems, with $\Lambda_1 = \Lambda_2 = 127$, $246$, and
$383$, respectively and are scaled to a total detector frame mass of $70M_\odot$ with a distance of $400$~Mpc. The 2B equation of state is almost identical to an $n = 0.5$ polytrope, only differing in the use of a realistic equation of state for the crust~\cite{Read:2009yp}.\footnote{There is a typo in Table~I of~\cite{Chen:2020fzm}---it should read $\Gamma_1 = 3$ for the 2B case.}

We find that for all of these simulated observations, there are cutoff frequencies and polytropic indices for which more than $90\%$ of the samples satisfy the contact frequency greater than cutoff frequency criterion. In addition, for the larger polytropic indices, at least one of the tidal deformabilities is below the minimum allowed for that polytropic index. We comment below on the results when computing the contact frequency using the black hole values for those samples, like above.

\begin{figure*}[tb]
	\centering
	\begin{tabular}{ccc}
	IMRPhenomD + PN tides simulated observation & \qquad\qquad\quad & scaled BNS simulated observation\\
	\includegraphics[width=0.45\textwidth]{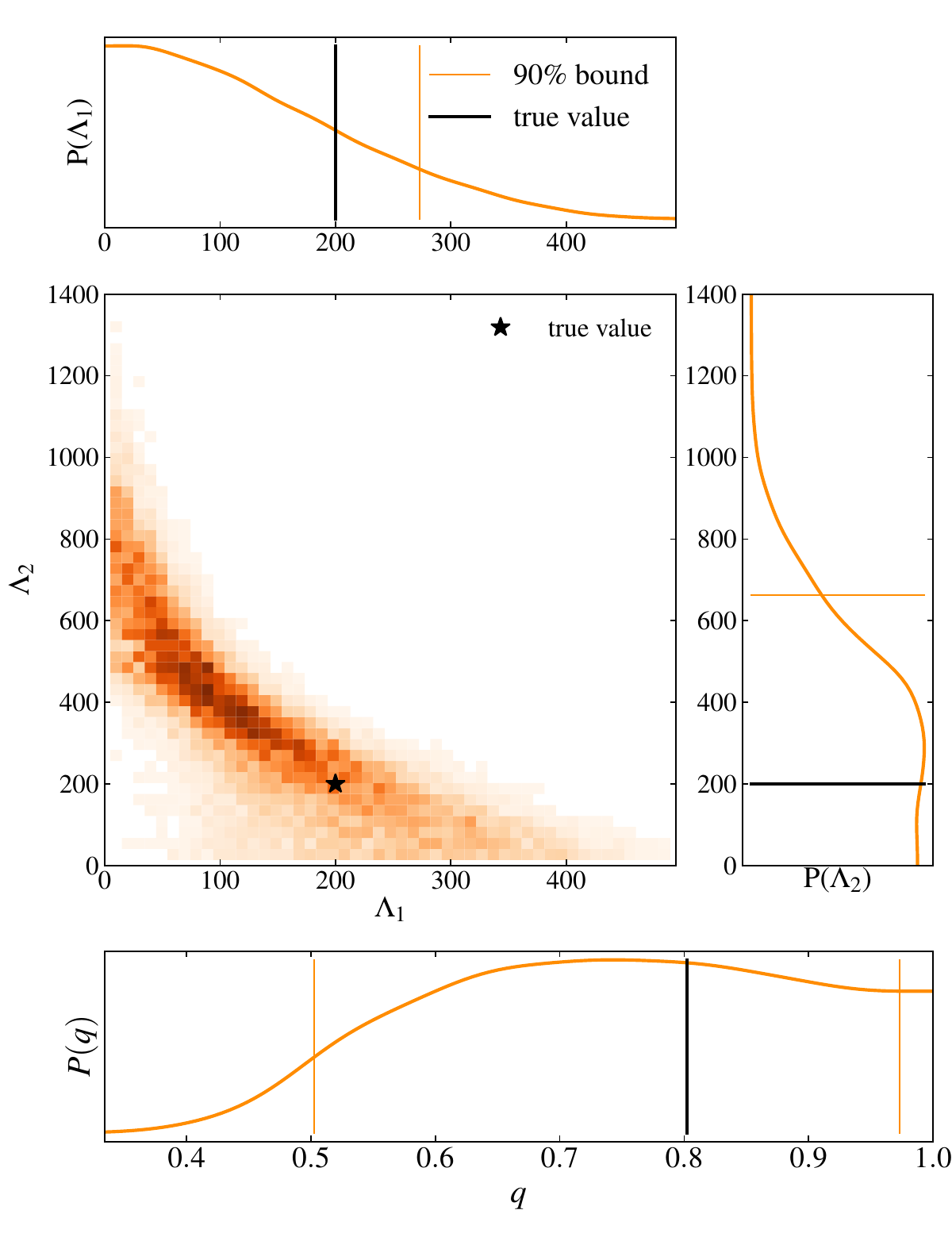} & &
	\includegraphics[width=0.45\textwidth]{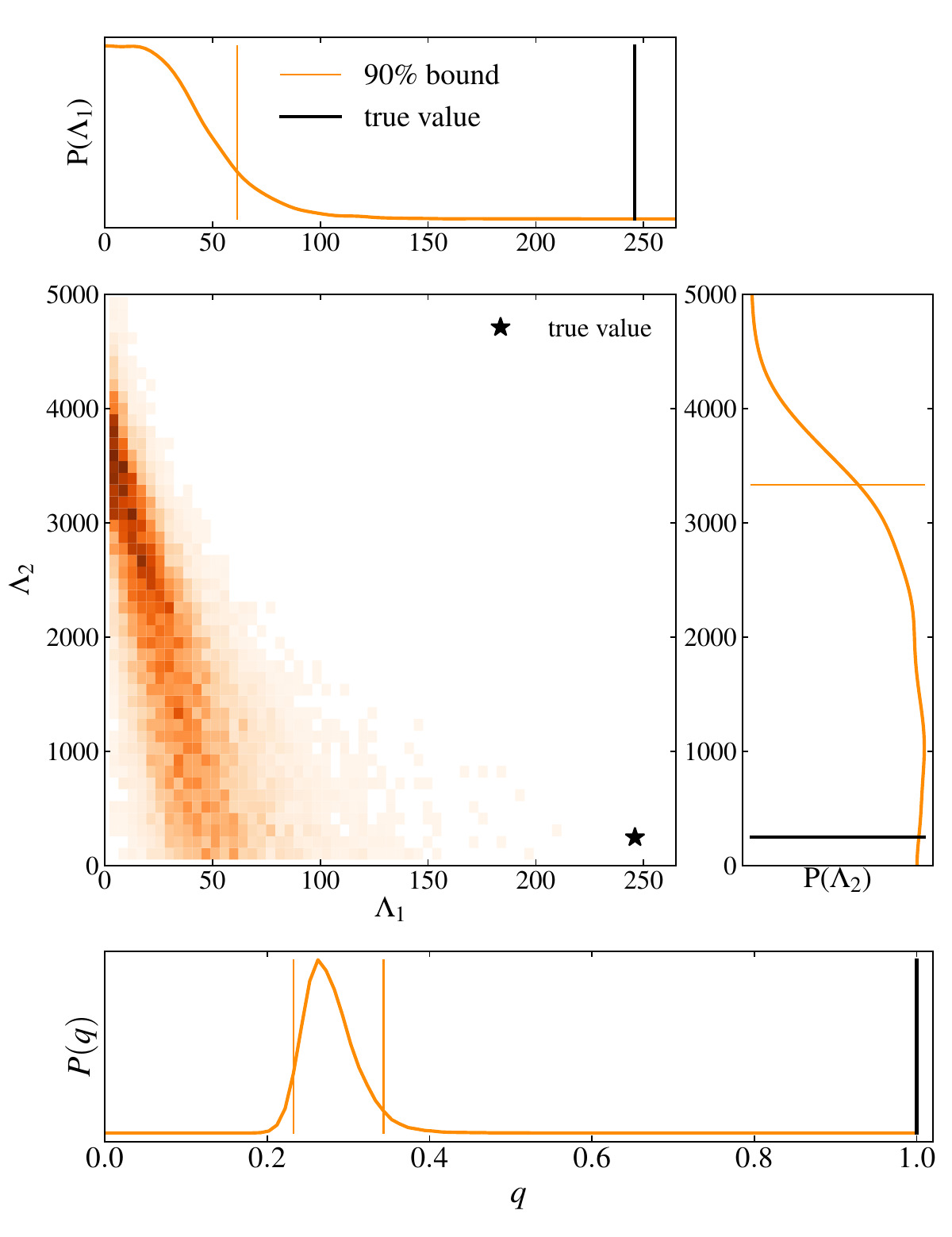}\\
	\includegraphics[width=0.45\textwidth]{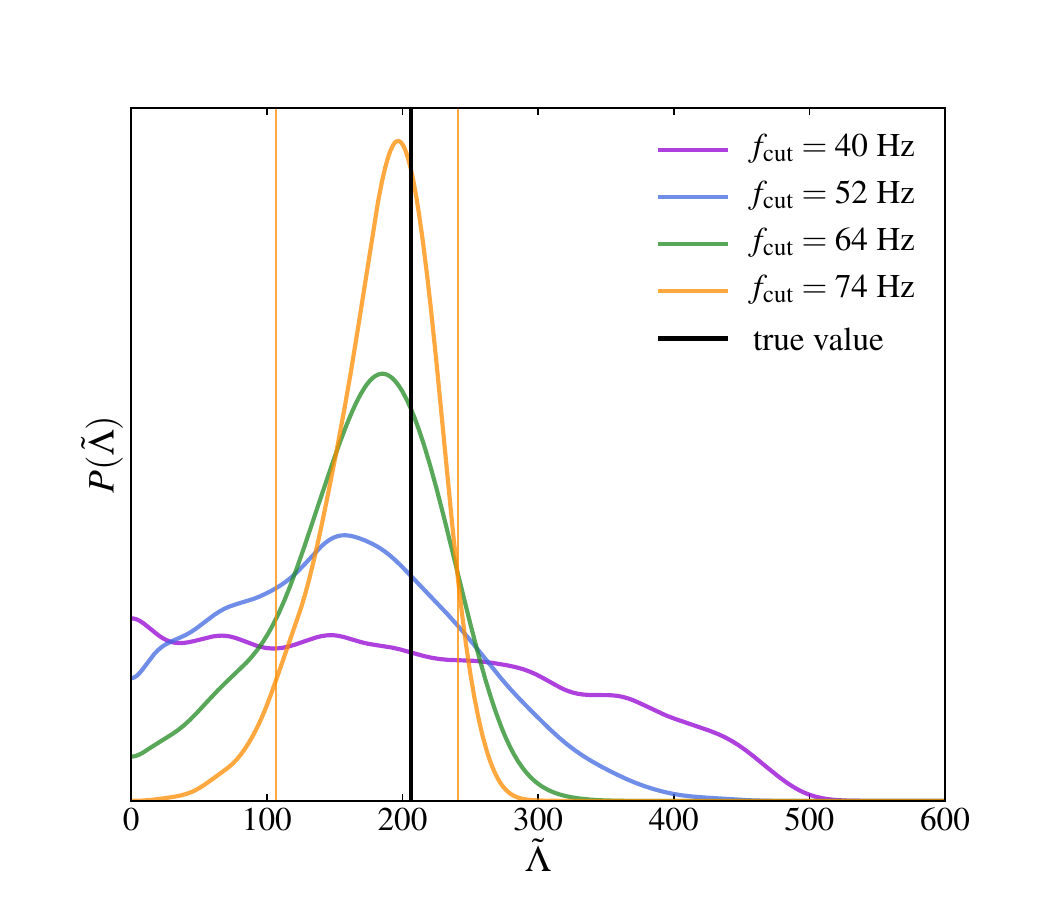} & &
	\includegraphics[width=0.45\textwidth]{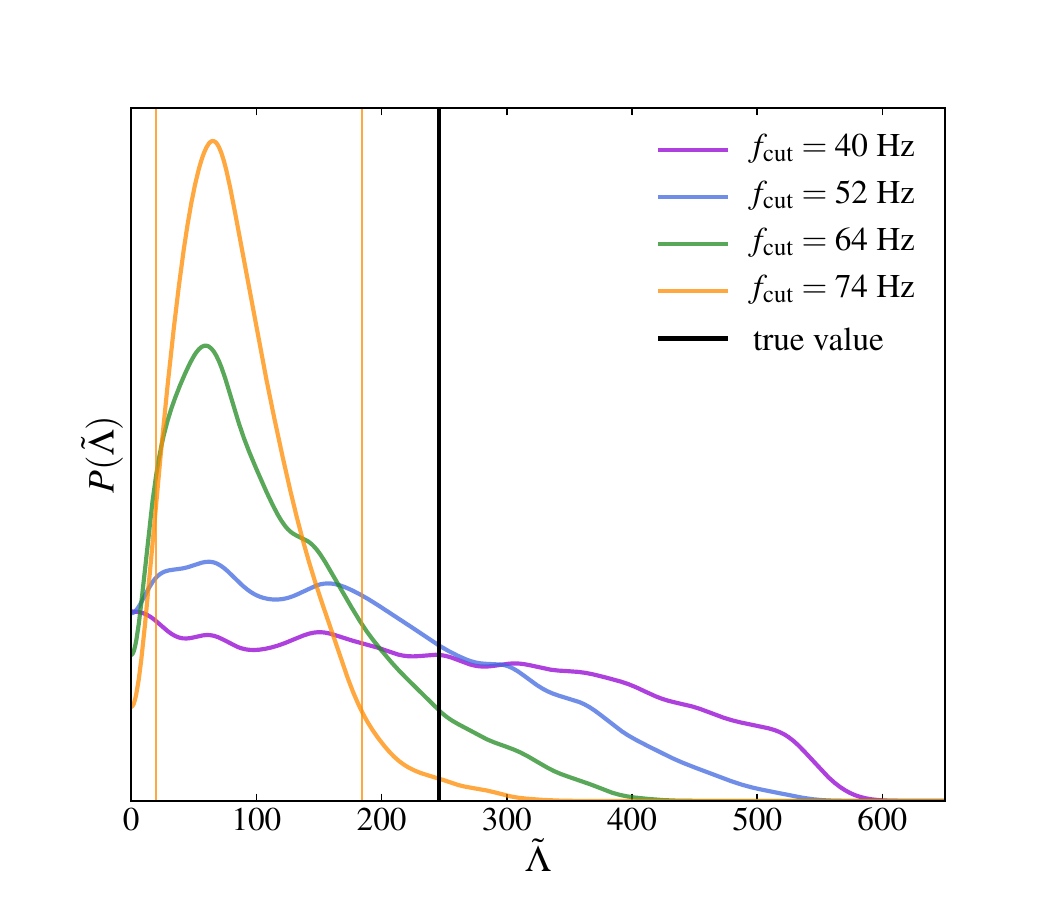}
	\end{tabular}
	\caption{\emph{Top panels:} Posterior distributions of the tidal deformabilities of the two objects, $\Lambda_1$ and $\Lambda_2$
(marginalized over all other parameters), obtained from simulated observations of two binaries with nonzero tidal
deformabilities, one \emph{(left)} with GW150914-like masses and distance and $\Lambda_1 = \Lambda_2 = 200$ created
with the same waveform model used in this analysis and one \emph{(right)} that is equal-mass with $\Lambda_1 = \Lambda_2 = 246$
and obtained by scaling a numerical binary neutron star waveform to a total mass similar to GW150914. These results are both
obtained with a cutoff frequency of $74$~Hz, the highest frequency we use in our analysis of these simulated observations. In each set of figures, the center panel
shows the two-dimensional distribution of $\Lambda_1$ and $\Lambda_2$ (darker shading indicates larger probability density), while the panels to its top and right show the marginalized
one-dimensional posteriors for these quantities and the bottom panel shows this for the mass ratio $q$. For $\Lambda_{1,2}$ we show the $90\%$ upper bounds, while for $q$ we show
the $90\%$ credible interval around the median. We see significant waveform
systematics in the right-hand figure. \emph{Bottom panels:} The posterior distributions of the effective tidal deformability $\tilde{\Lambda}$ of the binary
with a variety of different cutoff frequencies, reweighted to a uniform prior in $\tilde{\Lambda}$, for the same two simulated observations as the upper plots.
The vertical lines show the $90\%$ bounds on $\tilde{\Lambda}$ for the case with the highest cutoff frequency, again illustrating the waveform
systematics observed for the scaled BNS simulated observation.
	}
	\label{fig:posterior_plots_nonzero_tides}
\end{figure*}

We find that the effective tidal deformability $\tilde{\Lambda}$~\cite{Wade:2014vqa} posterior peaks away from zero for all these simulated observations---reweighting to a flat prior in $\tilde{\Lambda}$ as in, e.g.,~\cite{GW170817,Chen:2020fzm}---and the posteriors peak more sharply as the upper cutoff frequency is increased. However, for the ALF2 cases, with their larger $\tilde{\Lambda}$ values, the reweighted $\tilde{\Lambda}$ posteriors for the largest cutoffs peak well below the true values and the mass ratio posterior peaks well away from $1$. Similar waveform systematics are found when analyzing these simulated observations with their original BNS masses and a waveform model that augments a binary black hole waveform with a simple BNS tidal phasing~\cite{Chen:2020fzm}. Moreover, we find that the individual tidal deformabilities are not recovered well in any of these cases---they always peak at zero. In the worst cases, the $90\%$ upper bound on $\Lambda_1$ is well below the true value. These effects are the most pronounced for the ALF2 case with the larger tidal deformability, where there are no samples with $\Lambda_1$ and mass ratio at the true values. For the simulated observations of the same waveform we use for recovery, we find that even though the tidal deformabilities peak at zero and the mass ratio peaks away from unity, the true values are still in the bulk of the probability distribution. All of this is illustrated in Fig.~\ref{fig:posterior_plots_nonzero_tides}.

The neutron star equations of state we consider are similar to polytropic indices of $0.5$ to $0.8$, e.g., comparing the $\Lambda_{\ell=2}$ versus $M_\star/M_\text{max}$ curves, as in Fig.~\ref{fig:polytropic_and_boson_star_Lambda_vs_M}. This similarity is almost exact in the 2B case, as discussed above, and is much rougher in the ALF2 case. However, for those polytropic indices, we find that more than $90\%$ of the samples have contact frequency greater than cutoff frequency for the smallest high-frequency cutoffs we consider, but that the waveform systematics are still significant in these cases, particularly for the lower-mass ALF2 case, with its larger tidal deformability. Thus, future work will be necessary to refine this method of accounting for waveform systematics. For instance, one can compute the cutoff frequency using the black hole values when a tidal deformability is below the minimum allowed for the polytropic index under consideration. This reduces the fraction of samples with contact frequency greater than cutoff frequency for many of the aforementioned cases exhibiting significant waveform systematics. Another simple possibility is to set the maximum allowed surface deformation $\Delta R/R$ to be less than $0.2$. If this is reduced to $0.1$, then the maximum fraction of samples satisfying our acceptance criterion is reduced below $90\%$ in all these cases, except for the lowest cutoff frequency we consider for both the 2B and larger mass ALF2 cases. In those cases the fraction is still slightly above $90\%$, and thus one would need to reduce the maximum $\Delta R/R$ a bit further to make this fraction below $90\%$. Of course, reducing the maximum $\Delta R/R$ to $0.1$ also reduces the fraction of samples passing the acceptance criteria in the analysis of binary black hole simulated observations below $90\%$ in many cases, as well. Future work will consider more simulated observations and how best to set the maximum $\Delta R/R$, as well as alternate ways of dealing with waveform systematics and improvements to the waveform model.

However, even with the present method, one can make an additional check using the reweighted posterior on $\tilde{\Lambda}$. If this peaks away from zero, then this is a clear indication that one is not observing a binary black hole. We indeed find that the reweighted $\tilde{\Lambda}$ posterior peaks at zero for the binary black hole simulated observations and away from zero for the non-binary black hole simulated observations.

\section{Constraints on boson star models}
\label{sec:boson_star_constraints}

Here we show how the constraints on the maximum mass and minimum tidal deformability of 
black hole mimickers obtained in our general polytropic star framework can be translated 
into constraints on boson star models. These constraints should just be taken of an illustration
of the method, not a forecast for future constraints. Boson stars are not exactly described
by a polytropic EOS, thus the frequency cutoffs used to obtain these constraints are not exactly
the ones one would obtain in applying the same procedure to more realistic boson star models.
In addition, our simulated observations lack of realistic noise and calibration uncertainties. However, we find that
the tidal deformability versus mass curves of compact boson stars are well approximated by those
of polytropic stars (see Fig.~\ref{fig:polytropic_and_boson_star_Lambda_vs_M}), so these results are likely close to those that would be obtained in an 
analysis designed specifically for boson stars.

As discussed earlier, boson stars are equilibrium configurations of a massive, complex 
scalar field $\phi$. We consider boson stars with a boson mass $m_B$ and a quartic 
self-interaction given by a potential $V(|\phi|^2) = m_B^2 |\phi|^2 + \lambda_B |\phi|^{4}/2$, 
where $\lambda_B$ is the coupling constant of the quartic self-interaction term. Such stars 
were first considered in~\cite{1986PhRvL..57.2485C}, and their tidal deformabilities are 
calculated in~\cite{2017PhRvD..95h4014C, Sennett:2017etc}. One can produce the observed 
masses of the objects in the binary for sufficiently small boson masses even in the 
\emph{free-field} case ($\lambda_B = 0$). Specifically, one needs 
$m_B \lesssim (10^{-10} \text{ eV})(M_\odot/M_\text{max})$; see, e.g.,~\cite{Sennett:2017etc}. 
Increasing the self-interaction increases the allowed boson mass, though $\Lambda > 287$
for the quartic potential, regardless of the strength of the coupling~\cite{Sennett_PC}. 
Thus, the tidal deformability constraints presented in Fig.~\ref{fig:polytrope_constraints} 
would rule out a binary of quartic-potential boson stars as the source for these signals at the 
$90\%$ credible level, except for the low-$n$ results for the two less massive cases. However, 
low-$n$ polytropic stars (with $n \lesssim 1.4$) have minimum tidal deformabilities smaller 
than the minimum allowed for quartic potential boson stars, and thus should not be considered 
for this analysis. Thus, binaries of quartic-potential boson stars (and also free-field 
boson stars) can be ruled out at the $90\%$ credible level as potential sources for all four 
simulated events. 

\begin{figure}[tb]
\includegraphics[width=0.5\textwidth]{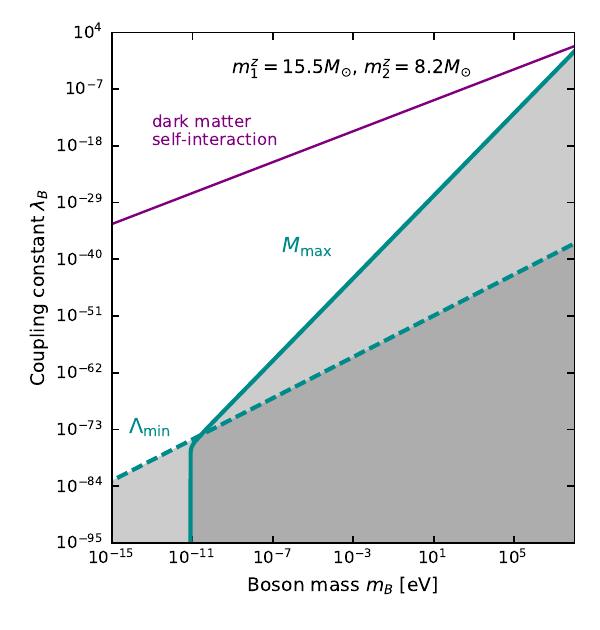} 
\caption{Constraints on the parameter space of a boson star with quartic self-interaction 
from the simulated observation of the $15.5, 8.2$ $M_\odot$ (detector frame) binary black hole using the fits in 
Eqs.~\eqref{eq:boson_star_fits} applied to the $n \geq 1.4$ results.  Here $m_{B}$ is 
the boson mass and $\lambda_B$ is the coupling strength. The solid and dashed cyan curves 
correspond to the observed lower limit of $m_{1} \geq 10.5 M_{\odot}$ and upper limit of 
$\Lambda \leq 298$ (both at the $95\%$ credible level), respectively. The gray region is 
excluded at the $95\%$ credible level. The purple line shows the parameters corresponding to 
a self-interaction cross-section per unit mass of $0.1$--$1$ cm${}^2$/g, as suggested for 
dark matter by various astronomical observations (discussed in, e.g.,~\cite{Giudice:2016zpa}).
}
\label{fig:int_mB}
\end{figure}

At higher credible levels, the upper bound on the tidal deformability is larger, and one only rules out some of the $(m_B, \lambda_B)$ parameter space with the simulated observations. We thus consider the $95\%$ credible level here, to illustrate these constraints on the parameter space, where we find that the least massive case can be produced by a binary of boson stars with a quartic self-interaction. (Note that $> 95\%$ of the samples have contact frequencies above the cutoff for the case we consider.)

The maximum mass and minimum tidal deformability of a stable boson star are given by~\cite{Sennett_PC}
\begin{subequations}\label{eq:boson_star_fits}
	\begin{align}\label{eq:Mmax_interacting_boson_star}
	M_\text{max} &\simeq 0.62 \, w^{-1/2} \, \frac{\MPl^2}{m_B}, \\
	\ln \Lambda_\text{min} &\simeq  1.706w^2 -1.198 w^{3/2} + 0.828w - 0.085w^{1/2} + 5.66,\label{eq:Lambda_min_fit}
	\end{align}
\end{subequations}
where $\MPl$ is the Planck mass and $w:= (1+ \MPl^2 \lambda_B/ 64\pi m_B^2)^{-1}$. These approximate expressions (with fractional accuracies of a few percent) are obtained from fits to the computations in~\cite{Sennett:2017etc}.

In Fig.~\ref{fig:int_mB}, we show the constraints on $m_B$ and $\lambda_B$ from the $95\%$ lower limit on the mass of the primary and $95\%$ upper limit on the tidal deformability for the simulated observation of the $15.5, 8.2$ $M_\odot$ (detector frame) binary black hole. For comparison, we also show the range of $(m_B, \lambda_B)$ values that corresponds to a self-interaction cross-section per unit mass of $\sigma_B/m_B = 0.1$--$1$ cm${}^2$/g. As discussed in~\cite{Giudice:2016zpa}, this range of self-interaction cross-sections is the order of magnitude required for dark matter to explain various observations. For the purposes of this illustration, we assume that the Lagrangian used to describe the boson stars describes all the interactions of the complex scalar field and use the tree level cross-section $\sigma_B = (3/32\pi)(\lambda_B/m_B)^2$ calculated in~\cite{Choi:2016hid}. The tree level approximation is very good everywhere except for the very highest boson masses we plot.

\section{Summary and outlook}
\label{sec:concl}

We have introduced a Bayesian method for constraining the properties of black hole mimickers 
such as boson stars or gravastars using GW observations. These constraints come from assuming 
that both members of the binary that generated the GWs belong to the same family of black hole 
mimickers, so that the measured masses and tidal deformabilities give a lower limit on the 
maximum mass and upper bound on the minimum tidal deformability allowed for this family. In 
the absence of accurate numerical calculations of waveforms from binaries of black hole 
mimickers, we model them as perfect fluid stars with a polytropic EOS and present a method to
account for waveform systematics. We have presented 
sample constraints on polytropic parameters using simulated observations of binary black 
holes with parameters similar to the first four binary black hole events observed by Advanced 
LIGO. We have also shown that extrapolating these constraints from
polytropic stars to boson stars rules out 
binaries of boson stars constructed using either a noninteracting scalar field or a quartically 
interacting scalar field model as sources of the modeled GW signals. However, these constraints
may be overly optimistic, as we have found that our method of addressing waveform systematics is not
quite adequate to prevent these effects from significantly affecting our results if the binary's constituents have
nonzero tidal deformabilities. Improving our method will be the subject of future work.

Additionally, our present work does not consider spin, which would introduce a number of complications, due 
to the presence of spin-induced deformations~\cite{Harry:2018hke} as well as spin-tidal 
couplings, which are just beginning to be explored (e.g.,~\cite{Pani:2015nua, Landry:2017piv, 
Abdelsalhin:2018reg, Landry:2018bil}). Improving the waveform models to include the spinning 
case will be the subject of future work, which will also use the higher-order nonspinning terms
recently computed in~\cite{Henry:2020ski}. However, for scalar boson stars, it may not be necessary to
include spin, as recent work~\cite{Sanchis-Gual:2019ljs,DiGiovanni:2020ror} has shown that spinning scalar boson stars
with no self-interaction or a quartic potential are unstable. Ultimately, one needs accurate numerical simulations of 
binaries of black hole mimickers with which to test current waveform models and calibrate new 
ones. Our use of polytropic stars also excludes cases where the tidal deformability 
is negative, e.g., most gravastar models~\cite{Pani:2015tga, Uchikata:2016qku}. We plan to 
consider gravastar models in future work.

As theoretical studies mature, and the sensitivity 
of advanced GW detectors increases, we are presented with a real possibility of significantly 
constraining theoretical models of exotic compact objects, or detecting such objects in nature.
Thus, in addition to numerical modeling of binaries of boson stars and other black hole mimickers, one
will need population synthesis calculations of such binaries, in order to constrain possible formation scenarios with
upper limits on (or measurements of) rates of black hole mimicker binary coalescences from gravitational
wave observations. Here improved experimental/observational constraints on the properties of dark matter
could play an important part in the theoretical modeling for black hole mimicker models related to dark matter.
One will also need further modeling of various models of black hole mimickers and their perturbations.
This is necessary both to model the gravitational wave signals of binaries as well as to allow one to discriminate
between various models (e.g., different potentials for boson stars) if binaries of black hole mimickers are detected,
by comparing the observed tidal deformability versus mass relation with the relation predicted by various models.\\

\section{Acknowledgments} 

We thank K.~G.~Arun, Tim Dietrich, Tanja Hinderer, Philippe Landry, Eric Poisson, and 
Rafael Porto for useful comments and discussion, Noah Sennett for providing clarifications 
about his work and producing the fit for the minimum tidal deformability, Tim Dietrich and Reetika Dudi
for the numerical simulation and creation of hybrid waveforms for binary neutron stars used here, An Chen
and Aditya Vijaykumar for assistance with the analysis of simulated observations with nonzero tides, and Aaron Zimmerman for a careful
reading of the manuscript. N.~K.~J.-M.\ acknowledges support from the AIRBUS 
Group Corporate Foundation through a chair in ``Mathematics of Complex Systems'' at the 
International Centre for Theoretical Sciences (ICTS) and from STFC Consolidator Grant 
No.~ST/L000636/1. Also, this work has received funding from the European Union's Horizon 
2020 research and innovation programme under the Marie Sk{\l}odowska-Curie grant agreement 
No.~690904. This research was supported in part by Perimeter Institute for Theoretical 
Physics. Research at Perimeter Institute is supported by the Government of Canada through 
Industry Canada and by the Province of Ontario through the Ministry of Economic Development 
\& Innovation. A.~M.\ acknowledges support by a ``Start-up grant for Young Scientists'' 
(SB/FTP/PS-067/2014) by the Science and Engineering Research Board (SERB), India and 
by the Science and Technology Research Council (Grant No.\ ST/L000946/1). P.~A.'s 
research was supported by a Ramanujan Fellowship from the SERB, by the Max Planck Society 
through a Max Planck Partner Group at ICTS and by the Indo-US Centre for the Exploration 
of Extreme Gravity funded by the Indo-US Science and Technology Forum (IUSSTF/JC-029/2016). 
LIGO was constructed by the California Institute of Technology and Massachusetts Institute 
of Technology with funding from the National Science Foundation and operates under cooperative 
agreement PHY-0757058. Computations were performed at the ICTS cluster Alice. This is LIGO 
document P1800092.

\bibliography{BHMimickers}

\end{document}